\documentclass[11pt,twoside]{article}
\usepackage{asp2010}

\resetcounters
\begin{document}

\title{pcigale: porting \emph{Code Investigating Galaxy Emission} to Python}
\author{Yannick Roehlly$^1$, Denis Burgarella$^1$, V\'eronique Buat$^1$, M\'ed\'eric Boquien$^1$, Laure Ciesla$^1$$^2$ and S\'ebastien Heinis$^1$$^3$}
\affil{$^1$Aix Marseille Universit\'e, CNRS, LAM (Laboratoire d'Astrophysique de
Marseille) UMR 7326, 13388, Marseille, France}
\affil{$^2$University of Crete, Department of Physics, Heraklion, Crete, 71003, Greece}
\affil{$^3$Department of Astronomy, CSS Bldg., Rm. 1204, Stadium Dr., University of Maryland, College Park, MD 20742-2421}

\begin{abstract}
We present pcigale, the port to Python of CIGALE (\emph{Code Investigating Galaxy Emission}) a Fortran spectral energy distribution (SED) fitting code developed at the \emph{Laboratoire d'Astrophysique de Marseille}. After recalling the specifics of the SED fitting method, we show the gains in modularity and versatility offered by Python, as well as the drawbacks compared to the compiled code.
\end{abstract}

\section{The origins}

CIGALE \citep{2005MNRAS.360.1413B, 2009A&A...507.1793N, roehlly2012CIGCodInvGALEmi} is a Fortran SED fitting code developed at the \emph{Laboratoire d'Astrophysique de Marseille}. From various models (single stellar populations, star formation history, attenuation law, infra-red templates\ldots) associated to free parameters, CIGALE builds theoretical spectral energy distributions (SEDs) and compares them to multi-wavelength observations of galaxies in order to derive the physical parameters for large galaxy samples. After years of development, the CIGALE team decided that the code needed to evolve to circumvent some of its current drawbacks:

\begin{itemize}
 \item The addition of new models and templates to the SED creation process should be easy, and this process should be modularised to allow multiple combinations of models.
 \item The analysis method should be coded independently of the SED creation process to allow the development of other statistical analysis methods using the same SED creation modules.
 \item Alternative uses should also be possible, like generating theoretical SEDs with the objective of modelling galaxy evolution, independently of a fitting analysis.
\end{itemize}

With the participation of the \emph{Centre de donn\'eeS Astrophysiques de Marseille} (CeSAM), it was decided to write a new code in Python to take advantage of its object model and of the growing Python community in astrophysics. This new code will be named pcigale (/psi.gal/).

\begin{figure}[htp]
  \begin{center}
    \includegraphics[width=0.60\textwidth]{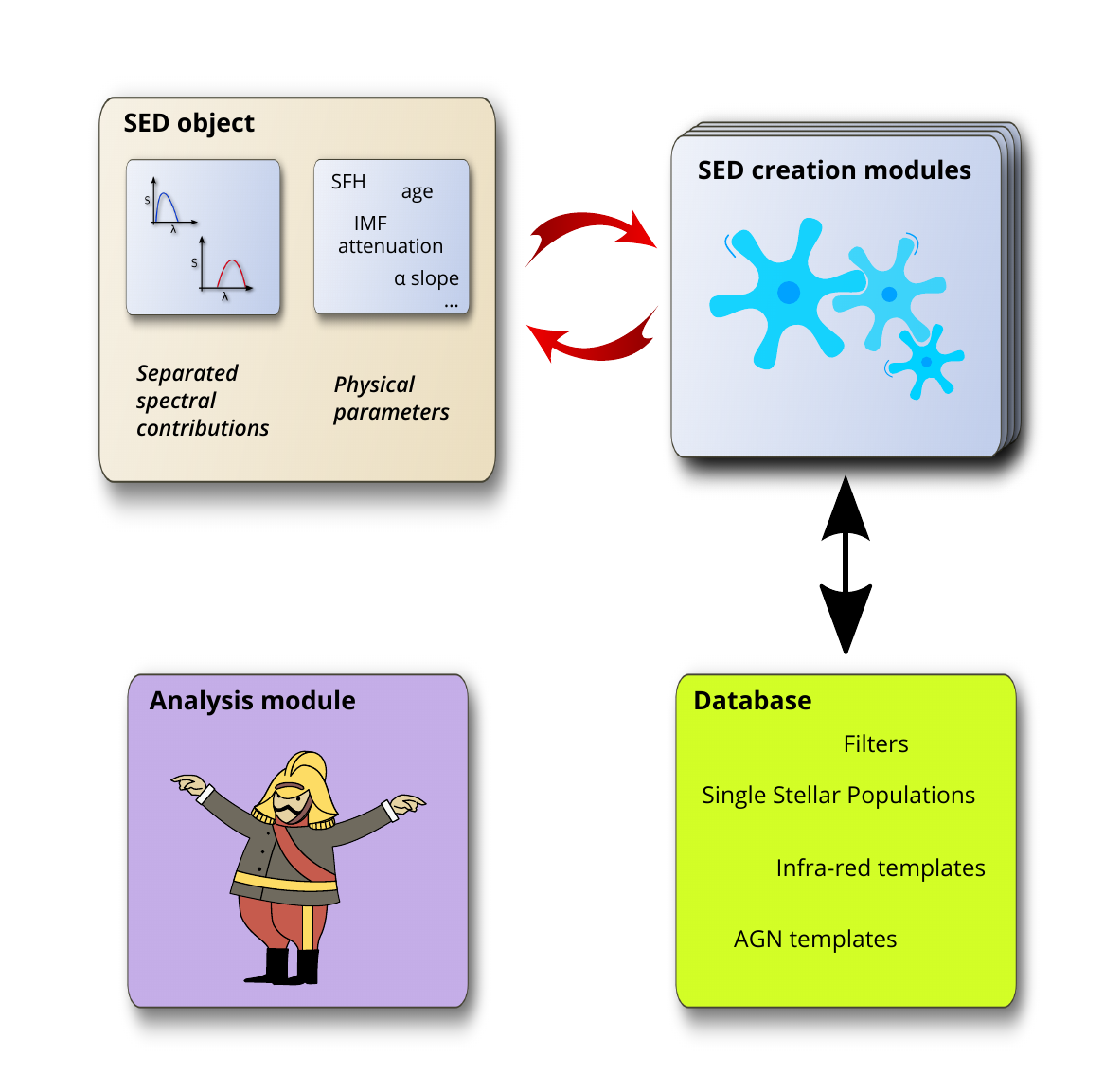}
    \caption{pcigale core organisation}
    \label{fig:organisation}
  \end{center}
\end{figure}

\section{pcigale core organisation}

pcigale is coded as a Python module rather than as a mere programme. The code is organised (see figure~\ref{fig:organisation}) around a \texttt{pcigale.sed} object with modules creating theoretical SEDs and other ones devoted to the analysis (\emph{e.g.} statistical fitting). 

\subsection{The SED object}

The \texttt{pcigale.sed} object keeps track of all the parameters associated with a SED, as well as of its \emph{history} (the module chain that led to its construction). The luminosity spectrum corresponding to this SED is stored as a list of cumulative sub-spectrum contributions that permits to identify the various components of a final SED.

\subsection{SED creation modules}

The SED creation modules are aimed at adding new information to a SED as well as some new components depending on specific parameters.

\begin{figure}[htp]
  \begin{center}
    \includegraphics[width=0.80\textwidth]{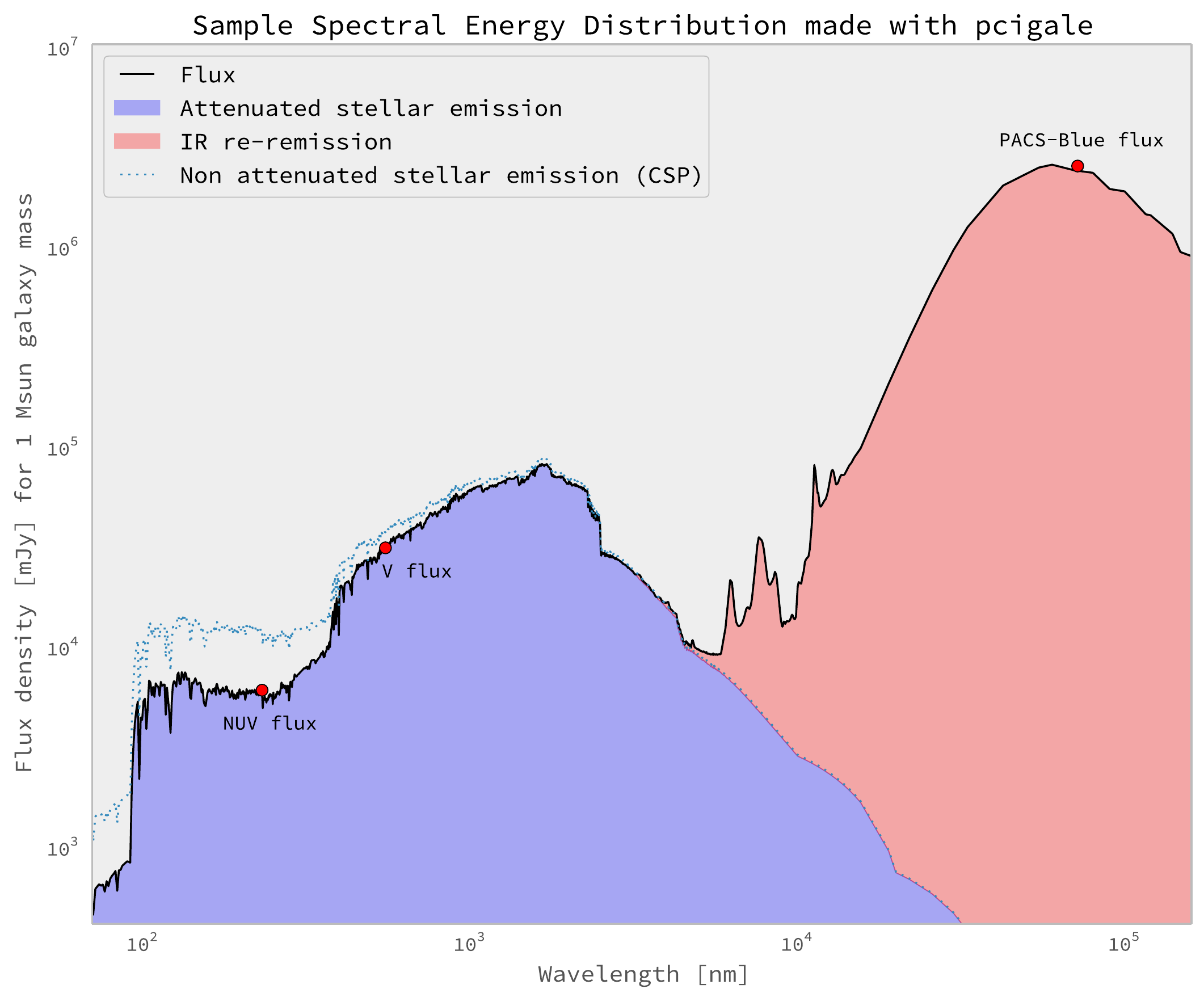}
    \caption{The various components of a constructed \texttt{pcigale.sed}.}
    \label{fig:samplesed}
  \end{center}
\end{figure}

For instance, a first module takes an empty \texttt{pcigale.sed} and adds a \emph{star formation history} (SFH, the evolution of the star formation rate along the lifetime of the galaxy) to it. A second module takes this SED and convolves its SFH with a \emph{Single Stellar Population} to add a stellar emission spectrum to the SED contribution list. A third module takes this spectrum, applies an \emph{attenuation law} to it and adds the corresponding (negative) contribution to the spectrum contribution list. A fourth module is used to determine the re-emission of the attenuated energy in \emph{infra-red} using templates. All these components are shown in figure~\ref{fig:samplesed}.

We made it easy to add new SED creation modules to pcigale. They are based on an abstract class and their code only needs to define their parameters and the way they process a \texttt{pcigale.sed} object. If needed, a database can be used to store the information useful for the processing (see figure~\ref{fig:organisation}).

\subsection{Analysis modules}

The analysis modules are the bandmasters which organise the specific use of the SED creation modules. For now, we wrote two modules: the first one performs a fitting analysis between models and photometric observations of galaxies, using a Bayesian-like method (similar to what the CIGALE original code does), the second module computes and saves theoretical SEDs in order to build artificial catalogues of galaxies. 

\subsection{A script}

For an easy pcigale use, we made a script that automates the analysis. Running pcigale to perform a SED fitting analysis reduces to the following instructions:
\begin{enumerate}
 \item \texttt{\$ pcigale init} -- a configuration file is created.
 \item Fill in the configuration file with the name of the flux table to be analysed, the list of modules used to create the SEDs and the name of the analysis module.
 \item \texttt{\$ pcigale genconf} -- new sections about the chosen modules are added to the configuration file.
 \item Fill in the configuration file again with the list of possible values for each parameters of the SED creation modules, as well as the analysis parameters.
 \item \texttt{\$ pcigale run}
\end{enumerate}

\section{Advantages and drawbacks of Python port}

The Python object model allows a good separation of concerns that permits the modularity described above and makes the code easily extensible. Compared to the Fortran version, we added new stellar population models, attenuation laws and infra-red and AGN templates.

Python is a multi-purpose language and benefits from a lot of modules. In the astrophysics field, \textbf{astropy} \citep{2013arXiv1307.6212T} makes it easy to deal with things like cosmology or reading and writing files. Last but not least, since the language is easy to learn, scientists with no specific programming skills can start to code rapidly.

The main drawback of Python is its speed compared to that of a compiled code. In particular, the \emph{first code} of a port will be slower than with the original Fortran code, which can be disappointing. To achieve speed gains, one might have to rewrite and optimise some parts of the code, for instance converting \texttt{for} loops in computation on Numpy arrays.

\section{Availability}

pcigale code will be released after a complete scientific validation. You can visit the CIGALE web page (\url{http://cigale.lam.fr}) or contact the team at \texttt{cigale@lam.fr}.


\bibliography{P87}

\end{document}